\documentclass[
  journal=nalefd,
  layout=twocolumn,
  manuscript=letter,
  doi=true,
  hyperref=true
]{achemso}

\usepackage{newtxtext,newtxmath} 

\usepackage{caption}
\usepackage{setspace}
\usepackage{titlesec}
\titlespacing*{\section}{0pt}{1ex}{0.5ex}
\titlespacing*{\subsection}{0pt}{0.8ex}{0.4ex}

\usepackage{chemformula} 
\usepackage[T1]{fontenc} 
\usepackage{enumitem}    
\usepackage{graphicx}    
\usepackage{braket}      
\usepackage{xcolor}      
\usepackage{amsmath}     

\makeatletter
\newlength{\@myfigwidth}
\if@twocolumn
\else
\fi
\newcommand{\myfigwidth}{\@myfigwidth}
\makeatother

\author{Michael Pfeufer}
\affiliation[JMU]
{Institute of Physical and Theoretical Chemistry, Julius-Maximilians-University Würzburg, Germany}
\author{Patrick Grenzer}
\affiliation[JMU]
{Institute of Physical and Theoretical Chemistry, Julius-Maximilians-University Würzburg, Germany}
\author{Friedrich Schöppler}
\affiliation[JMU]
{Institute of Physical and Theoretical Chemistry, Julius-Maximilians-University Würzburg, Germany}
\author{Tobias Hertel}
\affiliation[JMU]
{Institute of Physical and Theoretical Chemistry, Julius-Maximilians-University Würzburg, Germany}
\email{tobias.hertel@uni-wuerzburg.de}

\title[]{A Visible-Frequency Excitonic Reststrahlen Band in \(\mathrm{(PEA)_2PbI_4}\) Slabs}

\begin{document}

\begin{abstract}
Layered halide perovskites host exceptionally strong excitons, whose optical signatures are usually read as absorptive resonances on a smooth dielectric background. Strong excitons, however, can also reshape the dielectric response itself and drive the real permittivity negative, opening a reflective band---the visible, excitonic analogue of an infrared Reststrahlen band. Whether bare \(\mathrm{(PEA)_2PbI_4}\) slabs reach this regime has remained unclear. Here we show that low-temperature transmission of \(\mathrm{(PEA)_2PbI_4}\) slabs, driven by the intralayer-exciton manifold, evolves with increasing thickness from an excitonic dip into a broad near-zero-transmission interval with compressed Fabry--P'erot-like fringes. Transfer-matrix analysis with an effective Lorentz-oscillator dielectric response reproduces this crossover and reconstructs a finite negative-\(\mathrm{Re}(\varepsilon)\) window, and implies near-ultrastrong exciton--photon coupling. Calculated field maps show suppressed in-plane field penetration within this interval and a driven longitudinal response near the high-energy (\(\varepsilon=0\)) edge. These results identify \(\mathrm{(PEA)_2PbI_4}\) slabs as a cavity-free visible-frequency excitonic Reststrahlen material.
\end{abstract}

\section{Introduction}

Two-dimensional lead--halide perovskites are natural multiple-quantum-well semiconductors in which atomically thin inorganic sheets are separated by organic spacer layers.\cite{Ishihara90, Tanaka05, Katan19} In the (n=1) limit, quantum and dielectric confinement produce tightly bound excitons with large oscillator strengths,\cite{Blancon18, Katan19, Mauck19} while the layered crystal structure gives rise to pronounced optical anisotropy.\cite{Steger2022} Together, these properties make layered perovskites an attractive setting for studying strong light--matter interactions in a chemically tunable crystalline platform.\cite{Tamaki08, Yamamoto12, Straus18, Mauck19} Among them, phenethylammonium lead iodide, \(\mathrm{(PEA)_2PbI_4}\) (PEPI), has become an archetype of this family because its band-edge photoluminescence and absorption are dominated by strong excitonic resonances that persist as the defining spectral features from cryogenic to room temperature.\cite{Gauthron10, Steger2022, Grenzer2026} These same oscillator strengths also make the optical response of layered perovskites sensitive to exciton--photon hybridization once the electromagnetic boundary conditions allow optical modes to form.

This sensitivity has motivated extensive work on exciton--photon coupling, mostly in engineered photonic environments such as microcavities, plasmonic structures, and metasurfaces, where large normal-mode splittings and room-temperature polaritons have been demonstrated.\cite{Wang2018,Laitz2023} More recently, sufficiently thick layered-perovskite crystals have been shown to support cavity-free or self-hybridized exciton--polariton states, in which the crystal itself provides the photonic modes that hybridize with the excitons and the slab acts as a natural polaritonic medium.\cite{Anantharaman21, Black2024}

These studies frame the slab response in terms of Fabry--P\'erot-like photonic modes that hybridize with the exciton, yielding dispersive polariton branches with characteristic normal-mode splittings. What remains open is whether the exciton oscillator strength in a bare layered perovskite is sufficient to drive \(\mathrm{Re}(\varepsilon)\) negative---producing a Reststrahlen-like stop band in the visible, where reflectivity arises from the intrinsic dielectric response rather than from confined photonic modes alone.

PEPI is the obvious system in which to pose this question, since its exciton resonance is among the strongest and best characterized in the family. Yet the electrodynamic perspective has remained largely absent here, and its band-edge optical response has instead been discussed mainly in terms of exciton fine structure, dark states, out-of-plane transitions, and phonon- or polaron-dressed sidebands.\cite{Posmyk22, Posmyk24a, Posmyk23Phonon, Posmyk24b, Dyksik24Polaron} Accordingly, spectral shoulders and undulations around the main resonance have often been parsed into discrete excitonic features rather than into the electrodynamic response of a finite resonant slab. A recent combined experimental and theoretical study indeed shows that the low-temperature fine structure has a purely excitonic origin.\cite{Grenzer2026} What has remained unclear, however, is how to interpret the unusual reflectance and transmission spectra of thicker layered-perovskite slabs, and how these relate to a possible cavity-polaritonic origin of the near-zero-transmission window.

Resolving these questions requires modeling PEPI's optical response consistently across the full thickness range. Here we show that an intralayer-dominated effective Lorentz oscillator, propagated through the slab geometry by the transfer matrix, captures a crossover from a Lorentzian excitonic dip in thin films to a broad near-zero-transmission band with compressed low-energy Fabry--P\'erot fringes in thicker slabs. In the extracted dielectric response, the exciton manifold drives \(\mathrm{Re}(\varepsilon)\) negative over a finite spectral interval,\cite{Berreman63,Vassant12} producing a visible polaritonic stop band with a model-implied Rabi splitting in the ultrastrong-coupling regime. A mode and field analysis links this negative-permittivity window to slab exciton--polariton structure and to the excitonic photoluminescence response, establishing bare PEPI as a natural cavity-free exciton--polariton system.

\section{Results and Discussion}
\begin{figure*}[htbp]
    \centering
	\includegraphics[width=0.95\textwidth]{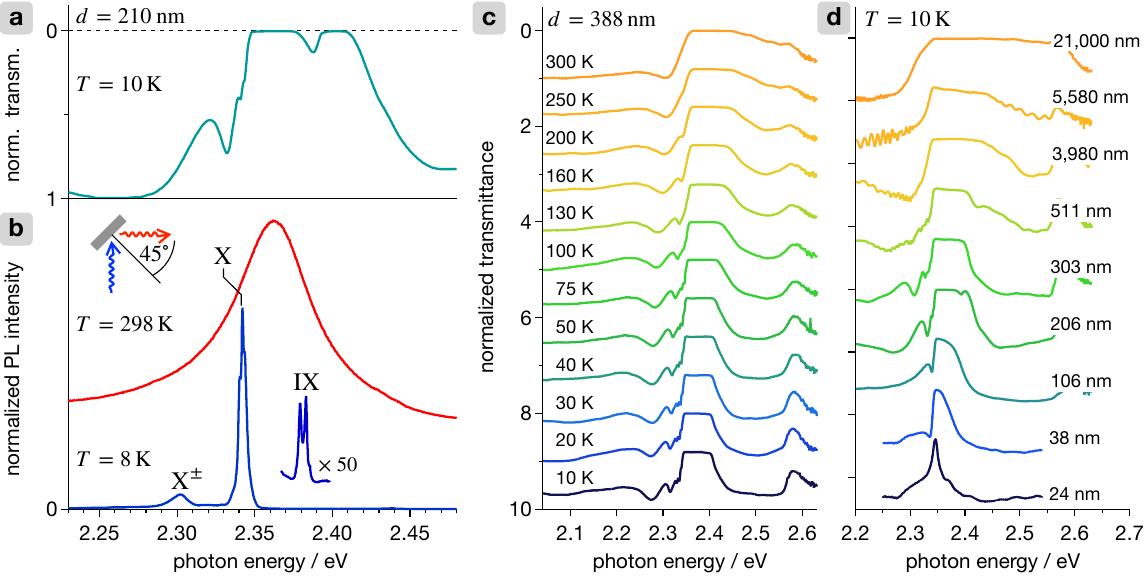}
    \caption{Excitonic energy scales and slab-transmission phenomenology of PEPI. (a) Inverted normalized transmittance spectrum of a \(d = 210\) nm slab at \(10\) K, showing a stop-band-like response near the intralayer exciton. (b) Room- and low-temperature PL spectra of a \(\sim 200\) nm slab, resolving at \(8\) K a trion-like feature near \(2.30\) eV and the intralayer-exciton manifold near \(2.34\) eV. Inset: expanded PL of a comparable flake showing a weak doublet near \(2.38\) eV, assigned in PL to interlayer excitons.\cite{Grenzer2026} (c) Temperature-dependent normalized transmittance of a \(d = 388\) nm slab from \(10\) to \(300\) K. (d) Thickness-dependent normalized transmittance at \(10\) K for slabs from \(24\) nm to \(21{,}000\) nm, showing the crossover from a Lorentzian-like excitonic dip to a broad near-zero-transmission band with low-energy Fabry--P\'erot fringes.}
    \label{fig1}
\end{figure*}

The low-temperature transmittance spectrum of a \(d=210\) nm PEPI slab in Figure~\ref{fig1}a provides the experimental starting point. Between \(2.35\) and \(2.41\) eV, the transmittance falls to a near-zero plateau. On the low-energy side, this is preceded by one pronounced and several weaker, more closely spaced undulations, while on the high-energy side, the transmittance starts to recover above \(\sim 2.41\) eV. Within the broad transmission minimum, a narrow revival is visible near \(2.38\) eV. Comparable spectra have previously been used to assist with the identification of exciton resonances within the PEPI optical response.\cite{Gauthron10,Do20b,Posmyk23,Posmyk24a} However, transmittance features do not map cleanly onto the different exciton manifolds resolved in the low-temperature PL spectrum of Figure~\ref{fig1}b.\cite{Grenzer2026} This mismatch suggests that the optical response of a finite-thickness PEPI slab is shaped by propagation effects that are typically negligible in the spectral analysis of atomically thin semiconductors.\cite{Mak2010,Li2014,Chernikov2014} In the following, we examine whether the intralayer-exciton manifold is sufficiently strong to produce a visible stop-band response in the slab transmittance, in analogy to the infrared Reststrahlen bands associated with LO--TO phonon-polariton resonances in ionic crystals.\cite{Caldwell15,Ma22}

The PL spectra in Figure~\ref{fig1}b provide the excitonic energy scale for this investigation. At low temperature, the PL resolves a weak trion-like \(\mathrm{X}^{\pm}\) feature near \(2.30\) eV, the bright intralayer-exciton manifold \(\mathrm{X}\) near \(2.34\) eV, and a very weak higher-energy doublet near \(2.38\) eV, recently assigned in PL to interlayer excitons in the two-layer unit cell of PEPI.\cite{Grenzer2026}

The temperature-dependent transmittance of a \(d=388\) nm slab in Figure~\ref{fig1}c shows that the near-zero transmission interval is a persistent feature of the slab optical response. At low temperature, the transmittance stays close to zero over a finite spectral range, with its low-energy onset broadly tracking the temperature-dependent upward shift of the \(\mathrm{X}\) manifold seen in PL \cite{Feldstein2020, Kaiser2021}. The compressed Fabry--P\'{e}rot fringes on the low-energy side of this range lose contrast with increasing temperature, consistent with thermal broadening of the excitonic response. A corresponding broadening is also evident in the room-temperature exciton PL in Figure~\ref{fig1}b. The plateau itself, however, remains clearly resolved up to at least \(160\) K and becomes only gradually less pronounced toward room temperature. A weaker dip near \(2.57\) eV is consistent with the onset of higher-energy band-to-band absorption but is secondary to our discussion of the exciton-driven response.

Figure~\ref{fig1}d shows how the low-temperature transmittance evolves systematically with slab thickness. For the thinnest flakes, exemplified by a \(24\) nm slab, the spectrum is dominated by a single minimum at the position of the intralayer exciton manifold, consistent with the absorption-dominated response expected for a quasi-two-dimensional semiconductor in the thin-film limit. With increasing thickness, this minimum broadens asymmetrically toward higher energy and, for thicknesses around \(200\) nm, develops into a broad near-zero-transmission plateau. As the slab thickness approaches the wavelength scale, interference fringes emergy on the low energy side of the plateau, and their nonuniform fringe spacing indicates strong excitonic dispersion of the slab refractive index. In the thickest slabs, additional fringes extend far below the \(\mathrm{X}\) resonance, where the refractive index varies more slowly and the response approaches that of an ordinary dielectric cavity. Optical thickness therefore governs the crossover from an absorptive excitonic resonance to a propagation-dominated stop band, where transmission through the slab is strongly suppressed over a finite spectral range. In the following, we use the term \textit{excitonic stop band} for this finite-thickness slab regime and, after reconstructing the negative-permittivity window, \textit{excitonic Reststrahlen band} to emphasize the analogy to phonon-polaritonic Reststrahlen bands in ionic crystals.

\begin{figure}[htbp]
  \centering
  \includegraphics[width=\columnwidth]{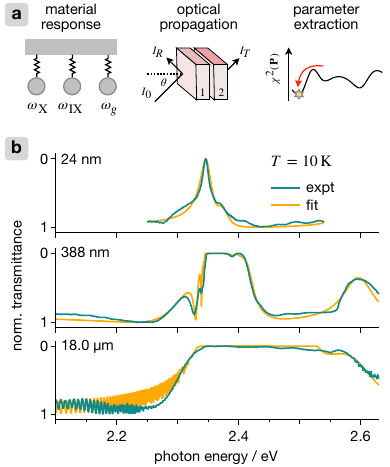}
  \caption{Transfer-matrix modeling of the thickness-dependent transmittance. 
  (a) Model workflow combining an effective Lorentz-oscillator dielectric response, transfer-matrix propagation through the air/PEPI/sapphire stack, and parameter extraction from measured spectra. (b) Normalized transmittance spectra and fits at \(T=10\) K for representative slab thicknesses of \(24\) nm, \(388\) nm, and \(18.0\,\mu\mathrm{m}\).}  
  \label{fig2}
\end{figure}

The thickness-dependent crossover motivates a transfer-matrix analysis of the slab transmittance. As summarized in Figure~\ref{fig2}a, the model combines an effective Lorentz-oscillator dielectric response with optical propagation through an air/PEPI/sapphire stack and fits the resulting spectra to the measured transmittance. In this way, a consistent material response can be propagated through different optical thicknesses and compared directly with the thin-, intermediate-, and thick-slab spectra.

The effective dielectric response was represented by the scalar Lorentz-oscillator form\cite{Yeh1988,BornWolf}
\begin{equation}
    \varepsilon(E)=\varepsilon_{\infty}+\sum_{j=1}^{3}\frac{A_j}{E_{0,j}^{2}-E^{2}-i\Gamma_j E},
\end{equation}
where \(\varepsilon_{\infty}\) is the background dielectric constant and \(A_j\), \(E_{0,j}\), and \(\Gamma_j\) are the oscillator strength, resonance energy, and damping of oscillator \(j\). The complex refractive index \(\tilde n(E)=n(E)+ik(E)=\sqrt{\varepsilon(E)}\) enters the transfer-matrix calculation for the air/PEPI/sapphire stack at \(45^\circ\) incidence. The dominant oscillator describes the intralayer exciton. A weaker interlayer-related term is included phenomenologically, motivated by the \(\mathrm{IX}\) emission feature in PL \cite{Grenzer2026} and by the small transmission revival occasionally observed within the suppressed region for slabs near \(200\) nm thickness. A third, higher-energy term captures the onset of band-to-band absorption. The oscillator parameters were obtained by minimizing the mismatch between calculated and measured transmittance spectra across the thickness series. Details of the optimization procedure and parameter constraints are provided in the Supporting Information.

Figure~\ref{fig2}b shows that this effective dielectric response reproduces the qualitatively different transmittance spectra observed across the thickness range. The Lorentzian-like excitonic minimum of the \(24\) nm slab, the broad near-zero-transmission plateau and compressed low-energy fringes of the \(388\) nm slab, and the dense fringe pattern extending far below the resonance in the \(18.0\,\mu\mathrm{m}\) slab are all captured within the same transfer-matrix description. The thickness-dependent crossover therefore follows from an intralayer-exciton-dominated dielectric response propagated through different optical thicknesses, rather than from distinct physical mechanisms in the thin-, intermediate-, and thick-slab regimes.

\begin{figure*}[htbp]
    \centering
	\includegraphics[width=0.95\textwidth]{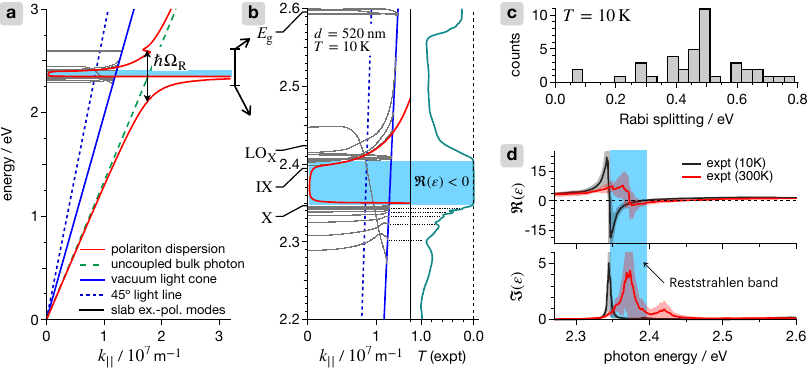}
    \caption{(a) Bulk-like polariton dispersion derived from the reconstructed dielectric response of a representative \(520\) nm slab at \(10\) K, together with the vacuum light cone (solid blue), the corresponding \(45^\circ\) light line (dashed blue), the uncoupled bulk-photon dispersion (green dashed), and the calculated finite-slab exciton--polariton branches (gray). The resulting Rabi splitting between upper and lower branches is indicated by \(\hbar\Omega_R\). (b) Mode structure calculated for the fitted \(d=520\) nm slab at \(10\) K, shown together with the corresponding experimental transmittance spectrum. The blue background marks the spectral interval with \(\mathrm{Re}(\varepsilon)<0\). Labels indicate the intralayer exciton (\(\mathrm{X}\)), interlayer exciton (\(IX\)), the model-derived longitudinal band-edge response (\(\mathrm{LO}_{\mathrm X}\)), and the band-edge energy (\(E_g\)). (c) Histogram of Rabi splittings extracted at \(10\) K from all analyzed slabs of different thicknesses. (d) Pointwise median of real and imaginary parts of the reconstructed dielectric function at \(10\) and \(300\) K. The blue shaded region again marks the range of negative \(\mathrm{Re}(\varepsilon)\) at 10 K, corresponding to the excitonic Reststrahlen band.}
    \label{fig3}
\end{figure*}

For each analyzed slab, the transfer-matrix fit yields an effective complex dielectric response, \(\varepsilon=\varepsilon_1+i\varepsilon_2\), from which a modal representation of the slab electrodynamics can be constructed. Figure~\ref{fig3}a shows the corresponding bulk-like polariton dispersion \(k(\omega)=\omega \tilde n(\omega)/c\) derived from the fitted dielectric function of a representative \(520\) nm slab at \(10\) K, together with the vacuum light cone and the associated finite-slab branches. Figure~\ref{fig3}b expands the energy scale around the stop band and excitonic manifold, and compares this p-polarized slab-mode structure with the measured transmittance, plotted vertically at the right side of the graph. At the \(45^\circ\) incidence used in the transfer-matrix fits, the incident field fixes a specific in-plane momentum \(k_\parallel\), corresponding to the dashed \(45^\circ\) vacuum light line shown in Figures~\ref{fig3}a,b. On the low-energy side of the stop band, the observed transmittance maxima track these resonant slab branches. Their compression toward the intralayer exciton reflects the strong curvature of the lower polariton branch in Figure~\ref{fig3}a as it approaches the excitonic resonance. In this representation, the light blue shaded interval with \(\mathrm{Re}(\varepsilon)<0\) corresponds to the excitonic stop band, where in-plane field penetration through the slab is strongly inhibited.

The same per-flake dispersion also provides a model-derived measure of exciton--photon hybridization in the slab. For each fitted \(10\) K spectrum, we determine \(\hbar\Omega_R\) from the minimum energy separation between the upper and lower \(k(\omega)\) branches, as illustrated for the \(520\) nm slab in Figure~\ref{fig3}a. Across the analyzed slabs, this procedure yields characteristic values around \(\hbar\Omega_R \approx 0.5\) eV, with the histogram in Figure~\ref{fig3}c showing the corresponding distribution. The resulting spread primarily reflects uncertainties in the transfer-matrix reconstruction. Relative to the intralayer-exciton energy \(E_{\rm X}=\hbar\omega_{\rm X}\approx 2.34\) eV, the corresponding ratio is \(\Omega_R/\omega_{\rm X}=\hbar\Omega_R/E_{\rm X}\approx 0.2\). Expressed in the common cavity-QED convention \(g/\omega_{\rm X}\), where \(\Omega_R\approx 2g\), this corresponds to \(g/\omega_{\rm X}\approx 0.1\), near the commonly used threshold for the ultrastrong-coupling regime.\cite{FornDiaz2019,Kockum2019}

The Rabi splitting defined in this way is conceptually distinct from the values commonly reported in layered-perovskite DBR microcavities, where the splitting quantifies the avoided crossing between a selected cavity mode and a single excitonic resonance.\cite{Wang2018,Laitz2023} The larger effective splitting obtained here reflects the collective oscillator response of the intralayer-exciton manifold throughout the optically active slab. It should therefore be understood as a model-derived splitting of the slab polariton dispersion, extracted from the reconstructed dielectric response, rather than as a direct experimental observable or a single-mode microcavity Rabi splitting.

\begin{figure*}[htbp]
    \centering
    \includegraphics[width=0.95\textwidth]{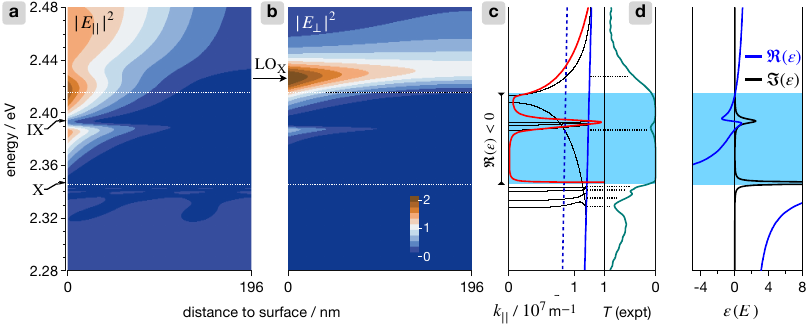}
    \caption{Calculated \(p\)-polarized field distributions, \(p\)-polarized slab-mode structure and dielectric response for a \(196\) nm PEPI slab on a substrate. Panels (a,b) show false color plots of the field intensities parallel and perpendicular to the slab surface, \(|E_\parallel|^2\) and \(|E_\perp|^2\), as functions of photon energy and distance from the air/slab interface. Panel (c) shows the corresponding slab-mode dispersion together with the experimental transmittance spectrum, using the same color code convention as in Figure~\ref{fig3}b. Within the negative-\(\mathrm{Re}(\varepsilon)\) stop-band region, \(|E_\parallel|^2\) is strongly suppressed inside the slab, whereas \(|E_\perp|^2\) develops a slab-spanning high-field feature near the upper \(\varepsilon=0\) crossing, labeled \(\mathrm{LO}_{\mathrm X}\) for its longitudinal, near-zero-permittivity character. Panel (d) shows the fitted dielectric response, \(\varepsilon(E)\), plotted on the same energy axis, with the interval \(\mathrm{Re}(\varepsilon)<0\) shaded in blue.}
    \label{fig4}
\end{figure*}
Figure~\ref{fig3}d summarizes the pointwise median effective dielectric functions obtained from the fitted spectra across the analyzed slabs. At \(10\) K, \(\mathrm{Im}(\varepsilon)\) shows a pronounced resonance at the intralayer exciton, while \(\mathrm{Re}(\varepsilon)\) crosses into a bounded negative-permittivity window over the same spectral range as the excitonic stop band. The shaded regions indicate the pointwise median absolute deviation (MAD) across the analyzed slabs. At \(300\) K, the excitonic resonance is broader and the dielectric response is corresondingly reduced. Nevertheless, \(\mathrm{Re}(\varepsilon)\) still crosses zero and remains negative over a narrower interval, consistent with the weakened but persistent stop-band structure at higher temperatures shown in Figure~\ref{fig1}c.

The combined evidence---a thickness-dependent near-zero-transmission interval, compressed low-energy fringes near the intralayer exciton, and a bounded negative-\(\mathrm{Re}(\varepsilon)\) window---supports identifying the stop-band regime in PEPI as an excitonic Reststrahlen band in the visible.\cite{Caldwell15,Passler17} We therefore analyze the internal field distribution in slabs near \(200\) nm thickness, where the optical response crosses over from a primarily excitonic transmission dip to a developed stop-band regime. Decomposing the field into components parallel and perpendicular to the slab surface reveals distinct field structures within and near the negative-\(\mathrm{Re}(\varepsilon)\) interval.

Figure~\ref{fig4} combines the calculated \(p\)-polarized field distribution at \(45^\circ\) incidence, the corresponding dielectric response, the \(p\)-polarized slab-mode dispersion, and the measured transmittance for a representative \(196\) nm slab. This allows the stop-band regime to be viewed simultaneously in real space, in \((E,k)\) space and in \(\varepsilon(E)\). We resolve the field inside the slab into components parallel and perpendicular to the slab surface, \(|E_\parallel|\) (in-plane, tangential) and \(|E_\perp|\) (out-of-plane, normal). The in-plane component \(|E_\parallel|^2\) captures the propagating, cavity-like field responsible for the Fabry--Pérot-type interference pattern, whereas the surface-normal component \(|E_\perp|^2\) highlights the longitudinal, near-zero-permittivity response near the high-energy edge of the negative-\(\mathrm{Re}(\varepsilon)\) window.

Within the stop-band region, marked by the blue shaded interval in Figures~\ref{fig4}c and d, \(|E_\parallel|^2\) shown in Figure~\ref{fig4}a is strongly suppressed inside the PEPI slab and decays rapidly away from the air/slab interface. This evanescent profile is the real-space representation of the stop-band response associated with the negative-\(\mathrm{Re}(\varepsilon)\) window. The Fabry--Pérot-like standing-wave pattern re-emerges on the low-energy side of the stop band, where \(\mathrm{Re}(\varepsilon)\) becomes positive again.

The normal field component shows complementary behavior near the high-energy edge of the stop band. As shown in Figure~\ref{fig4}b, \(|E_\perp|^2\) develops a pronounced high-field feature near \(2.43\) eV, about \(15\) meV above the upper end of the stop band, and extends through the entire slab thickness. The apparent field enhancement and its small offset are consistent with a longitudinal Berreman-like thin-film response:\cite{Berreman63} in a lossy finite slab at oblique incidence, the longitudinal-field maximum is governed by \(-\mathrm{Im}(1/\varepsilon)\) and by the full slab geometry, rather than by the bulk \(\mathrm{Re}(\varepsilon)=0\) condition alone. It can therefore be shifted by damping, slab thickness, substrate boundary conditions, and the finite in-plane momentum of the incident \(p\)-polarized field. Because this response appears most clearly in the calculated driven field distribution and not as a well-separated branch in the slab-mode dispersion, we use the label \(\mathrm{LO}_{\mathrm X}\) only as a descriptor for a model-derived longitudinal band-edge response, rather than as the name of an independently observed slab eigenmode.

A weaker field feature is also obtained for slabs with thicknesses near \(200\) nm that show faint internal structure within the stop-band region. Its energy overlaps with the weak interlayer-exciton doublet near \(2.38\) eV seen in the low-temperature PL of Figure~\ref{fig1}b, which has been associated with the layer inequivalence of the AB-type stacking order in PEPI.\cite{Grenzer2026} Within the effective dielectric model, this feature can be described as an interlayer-related (IX) perturbation of the longitudinal field response. Because it is weak and thickness-sensitive, however, we treat it as a secondary observation.

Transmission spectroscopy and transfer-matrix modeling show that PEPI slabs cross over from a Lorentzian-like excitonic thin-film response to a thickness-enabled stop-band regime with near-zero transmission and strongly compressed low-energy Fabry--Pérot-like fringes. The reconstructed dielectric response contains a finite negative-\(\mathrm{Re}(\varepsilon)\) interval over the same spectral range. The corresponding mode and field analysis further links this regime to slab exciton--polariton structure, suppressed in-plane field penetration, and a longitudinal band-edge response near the high-energy \(\varepsilon=0\) edge.

\section{Conclusion}

We have shown that \(\mathrm{(PEA)_2PbI_4}\) slabs exhibit a thickness-dependent crossover from a thin-film excitonic absorption feature to a broad visible-frequency stop-band response governed by the strong intralayer-exciton manifold, with the characteristic signatures of an excitonic Reststrahlen band in the visible.\cite{Caldwell15,Passler17} The measured transmission spectra are consistently described by a transfer-matrix model with an excitonic dielectric response that develops a finite interval of negative \(\mathrm{Re}(\varepsilon)\) at low temperature. From the same fitted dielectric response, the slab-polariton dispersion yields a model-derived Rabi splitting \(\hbar\Omega_R \approx 0.5\) eV, corresponding to \(g/\omega_{\mathrm X}\approx 0.1\) near the commonly used threshold for the ultrastrong-coupling regime. This splitting reflects the collective oscillator response of the intralayer-exciton manifold across the slab and is conceptually distinct from the single-mode Rabi splittings reported for layered-perovskite DBR microcavities.\cite{Kockum2019}

An important implication is that the pronounced spectral undulations observed in slabs with a clear stop-band signature arise from the slab electrodynamics of a strongly dispersive excitonic medium, including thickness-dependent interference, model-derived exciton--polariton mode structure, and the formation of the negative-permittivity stop-band regime. In this thickness range, they are therefore better viewed as slab-electrodynamic features than as a direct spectral map of the established layer-resolved excitonic fine structure \(\mathrm{D}_L,\mathrm{X}_L,\mathrm{Y}_L,\mathrm{Z}_L\) with \(L\in\{1,2\}\). This distinction matters more broadly for the interpretation of reflectance and transmittance spectra in layered perovskites, where slab-optical and polaritonic effects can mimic or obscure intrinsic excitonic fine structure.

The reconstructed dielectric response, modal analysis, and calculated field maps together indicate that the low-temperature stop-band regime is accompanied by suppressed in-plane field penetration and a driven normal field enhancement on the high-energy side of the excitonic stop-band region, analogous to Berreman-like thin-film physics near an \(\varepsilon \approx 0\) condition.\cite{Berreman63} This identifies PEPI as a chemically tunable layered semiconductor platform in which strong excitonic resonances directly reshape slab electrodynamics, including longitudinal band-edge responses, without the need for an external cavity.\cite{Anantharaman24}

Several questions remain open. Moving beyond the present isotropic description toward an anisotropic dielectric model will require angle- and polarization-resolved reflectance and spectroscopic ellipsometry on the same slab geometries,\cite{Steger2022,Song2020,Guo2018} clarifying how in-plane and out-of-plane oscillator strength, damping, and dispersion shape the stop-band response.\cite{Katan19,Li20} This direction is motivated by Steger \emph{et al.}, whose uniaxial analysis of PEPI resolved a weak out-of-plane exciton roughly \(40\,\mathrm{meV}\) above the intralayer manifold\cite{Steger2022}---a separation matching that between the intralayer and interlayer manifolds in the low-temperature PL of Figure~\ref{fig1}b\cite{Grenzer2026} and supporting its inclusion as an additional weak Lorentzian oscillator. A fully anisotropic model may thus refine the microscopic interpretation of the longitudinal band-edge enhancement identified here. A targeted thickness series near \(200\) nm would further clarify why weak interlayer-exciton structure appears most prominently in the crossover between the thin-film excitonic response and the developed stop-band regime.



\bibliography{references}
\end{document}